\documentclass[aps,prb,superscriptaddress]{revtex4}
\usepackage{graphicx}
\usepackage{bm}
\begin{document}
\title{Stability of zero-mode Landau levels in bilayer graphene against disorder 
in the presence of the  trigonal warping}

\author{Tohru Kawarabayashi}

\affiliation{Department of Physics, Toho University, Funabashi 274-8510, Japan}
\author{Yasuhiro Hatsugai}
\affiliation{Institute of Physics, University of Tsukuba, Tsukuba 305-8571, Japan}
\author{Hideo Aoki}
\affiliation{Department of Physics, University of Tokyo, Hongo, Tokyo 113-0033, Japan}

\begin{abstract}
The stability of the zero-energy Landau levels
in bilayer graphene against the chiral  symmetric disorder is examined in the presence of the
trigonal warping. Based on the tight-binding lattice model with a bond disorder correlated over 
several lattice constants, it is shown that among the four 
Landau levels per spin and per valley, two Landau levels exhibit the anomalous sharpness 
as in the absence of the trigonal warping, while the other two are 
broadened, yielding split peaks in the density of states. This can be attributed to the fact that 
the total chirality in each valley is $\pm 2$, which is protected topologically even in the 
presence of an intra-valley scattering due to disorder.
\end{abstract}

\maketitle

\section{Introduction}
In bilayer graphene, the Landau level structure in the 
vicinity of the contact point  is drastically changed 
by the trigonal warping. In the presence of the trigonal warping, the four Dirac cones 
appear in low energies out of the parabolic bands touching at K and K' points \cite{MF}. 
Due to this 
modification, we have four-fold zero-energy Landau levels 
per spin and per valley in weak magnetic fields. 
So the  trigonal warping doubles the 
degeneracy of the zero-energy Landau levels
from the situation in the absence of the trigonal warping.  
Now an interesting question is how the four Landau levels 
react to disorder.

Bilayer graphene, in the absence of the trigonal warping, has two-fold degenerated zero-energy Landau levels, 
and the stability of such zero-modes against disorder in gauge 
degrees of freedom (ripples) has been shown to be 
a consequence of the index theorem \cite{KP}.
We have also demonstrated with an explicit numerical approach based on a tight-binding model as well as  with 
Aharonov-Casher argument that the doubly-degenerated zero-energy Landau levels ($n=0$ and $n=1$) exhibit 
an anomalous sharpness for a bond disorder spatially correlated over more than 
several lattice constants and respecting the chiral symmetry of the system \cite{KHA_BG}. 
So the question now is whether the additional two zero-energy Landau levels in the presence of the trigonal warping in a weak magnetic field 
are all robust against  the bond disorder as well.
It is to be noted that the chiral symmetry itself is respected by the trigonal warping.

To clarify the stability of these zero-energy Landau levels  in the presence of 
the trigonal warping,  we have performed 
numerical calculations with a tight-binding lattice model in weak magnetic fields to examine the robustness against the 
spatially-correlated bond disorder.   
We shall find that, among the four Landau levels per spin and per valley, two levels exhibit the 
anomalous sharpness against  the finite-ranged bond disorder as in the absence of the trigonal warping, while the other two are broadened. This can be attributed to the fact that  the {\it total chirality of the  four Dirac cones}  
at K(K') is $+2(-2)$, and that, remarkably enough,  these are topologically protected even in the presence of the intra-valley scattering arising from the disorder.

\section{Model}
\begin{figure}
\begin{center}
\includegraphics[width = 1.0 \textwidth]{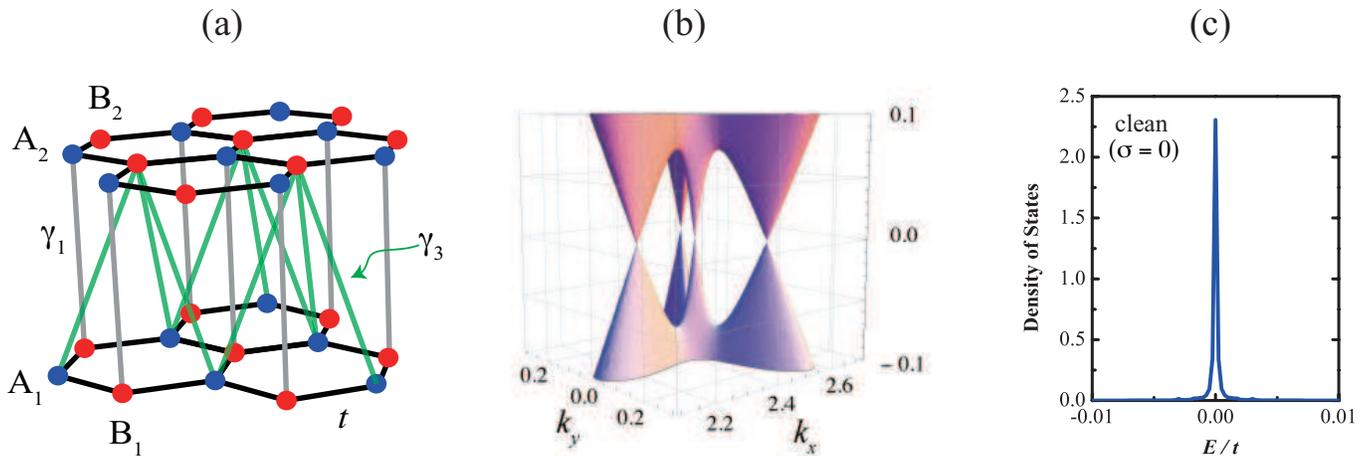}
\end{center}
\caption{\label{fig1}(a)  Lattice model for bilayer graphene with the largest 
interlayer coupling $\gamma_1$ and another, $\gamma_3$, that gives 
rise to the trigonal warping. 
(b) Energy dispersion $E(k_x,k_y)/t$ around  $(k_x,k_y) = (4\pi/3\sqrt{3},0)$ for the interlayer coupling 
$\gamma_1/t=0.2$ and $\gamma_3/t = 1.5$. (c) Density of states around $E=0$ for $\phi/(h/e) = 1/5000$ evaluated by the Green function $G(E+i\varepsilon)$
with $\varepsilon /t = 1.0 \times 10^{-4}$, which can be fit as 
$(4\phi/\pi\phi_0)(\varepsilon/(E^2+\varepsilon^2))$.}
\end{figure}

We adopt a tight-binding  lattice model for bilayer graphene with the Bernal (A-B) stacking having the inter-layer
couplings $\gamma_1$ and $\gamma_3$ (Fig.\ref{fig1}(a)).  The coupling $\gamma_1$ connecting 
the A$_2$ site in the top layer and the B$_1$ in the bottom  determines the overall parabolic dispersion, while the 
coupling $\gamma_3$ between  
the B$_2$ and the A$_1$ introduces the trigonal warping of the 
Fermi surface and produces the four Dirac cones at K(K') point in a low-energy scale \cite{MF}.
We introduce the bond disorder $\delta t$ in the nearest-neighbor hopping $t$ in each layer as in 
Ref.\cite{KHA_BG}.  The disorder is assumed to have a gaussian distribution with variance $\sigma$ and have a spatial correlation as $\langle 
\delta t(\bm{r}) \delta t(\bm{r}') \rangle = \sigma^2 \exp (-|\bm{r} - \bm{r}'|^2/4\eta^2)$ with the correlation length $\eta$.  In order to access to 
the low-energy regime in weak magnetic fields, we consider a system 
as large as $4\times 10^6$ sites
which is 4 times larger than those in our previous studies \cite{KHA_BG,KHA_MG} 
to evaluate the density of states from the Green function $\rho = -{\rm Im}\langle G_{r,r}(E+i\varepsilon) \rangle_r/\pi$
\cite{SKM}.  The magnetic field applied 
perpendicular to the graphene sheet is taken into account by the Peierls phases for the hopping amplitude, 
where the magnetic flux per hexagon of the honeycomb lattice is denoted by $\phi$ in units of the flux quantum $\phi_0=h/e$.  
Spin degrees of freedom are neglected for simplicity, and all the 
lengths are measured in units of 
the nearest-neighbor distance, $a$, of the honeycomb lattice. 

\section{Numerical Results}
In order to study numerically the low-energy physics where the trigonal warping (coming from $\gamma_3$) is
relevant, we assume  $\gamma_1/t = 0.2$ and $\gamma_3/t = 1.5$ in the above lattice model. Although 
the present value of $\gamma_3$ is much larger than the value estimated for the bulk graphite $\gamma_3/t \sim 0.1$ \cite{DD}, 
the dispersion at low energies is still described by the four Dirac cones as shown in Fig.\ref{fig1}(b),  
which should be adiabatically connected to the realistic situation with $\gamma_1/t = 0.12$ and $\gamma_3 /t= 0.1$. 
It is expected therefore that the stability of the 
zero-mode Landau levels against the disorder scattering, which mixes the four Dirac cones, can be discussed 
within the present lattice model by assuming an appropriate (see below) value of the correlation length $\eta$ of the bond disorder.  

First, we show in Fig.\ref{fig1}(c) the density of states for $\phi/\phi_0= 1/5000$ in the absence of disorder ($\sigma =0$), in which we 
clearly see that the 
zero-energy Landau level is 4-fold degenerated per valley. It is to be noted that the zero-energy Landau levels 
in a tight-binding lattice model with a finite system-size have small but finite widths. Exact zero-energy states with the $\delta$-function density of states
are realized only in the thermodynamic and continuum limit. In the present calculation, such a small width, however, is 
much smaller than the energy resolution determined  by the imaginary 
part $\varepsilon$ of the energy, since the sharp peak in the absence of disorder can be fitted fairly accurately by 
the Lorentzian form $\varepsilon /(E^2+\varepsilon^2)$ with a pre-factor consistent with the four-fold degeneracy
of the Landau levels.

\begin{figure}
\begin{center}
\includegraphics[width = 1.0 \textwidth]{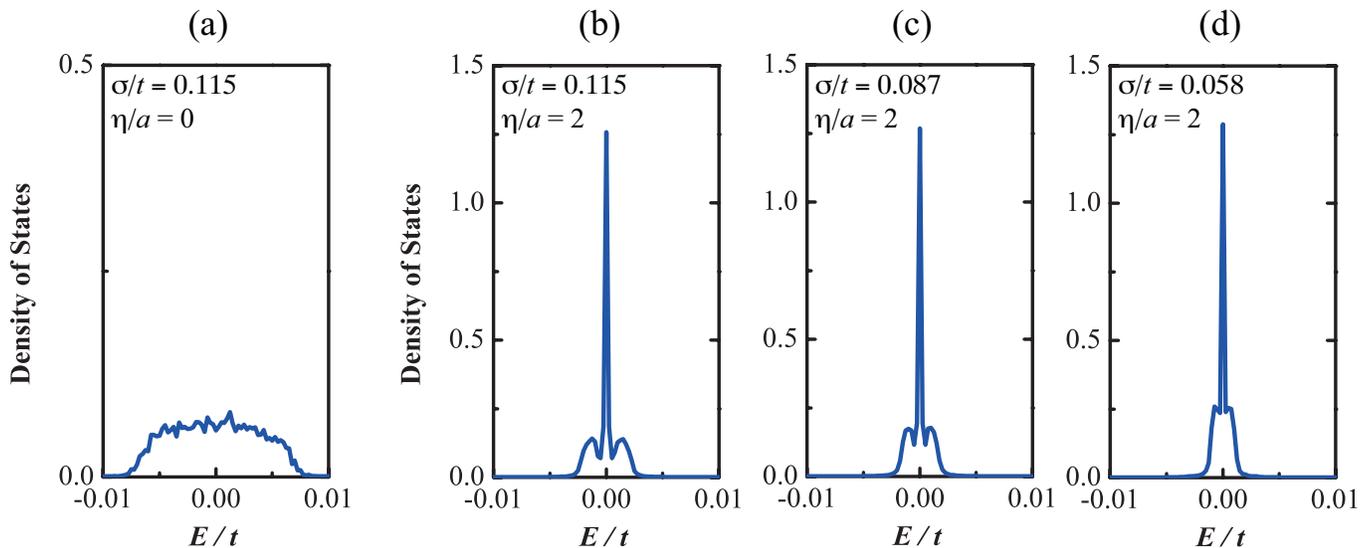}
\end{center}
\caption{\label{fig2} Density of states at zero energy for  an uncorrelated disorder with 
$\eta =0$ (a) and for a correlated disorder with $\eta /a = 2$ ((b),(c),(d)). 
Disorder strength is assumed to be $\sigma /t = 0.115$ for (a) and (b), while $\sigma /t = 0.087$ for (c) and 
$\sigma /t = 0.058$ for (d). The magnetic flux per hexagon is $\phi/\phi_0 = 1/5000$ and 
the imaginary part of energy is  $\varepsilon /t = 1.0 \times 10^{-4}$.}
\end{figure}

We then introduce the disorder in Fig.\ref{fig2}.  
For the case where the bond disorder is uncorrelated in space ($\eta =0$),   we confirm that the usual broadening occurs 
for the zero-energy Landau levels (Fig.\ref{fig2}(a)). In contrast, when the disorder is correlated over a few 
lattice constants ($\eta /a = 2$), we have one central sharp Landau 
levels accompanied by two broadened ones (Fig.\ref{fig2}(b),(c),(d)).
The peak height of the central peak is one-half of that in the absence of disorder (Fig.\ref{fig1}(c)), 
and its shape is insensitive to the disorder strength $\sigma$. The splitting of the broadened satellite peaks becomes larger for 
stronger disorder. The satellite peaks indicate the splitting of the critical energy for the 
quantum Hall transition, which 
has been discussed for the monolayer graphene with a  short-ranged bond disorder \cite{KA}.

All these results suggest that among four-fold zero energy Landau levels, two Landau levels remain anomalously sharp 
against the spatially correlated bond disorder as in the absence of the trigonal warping ($\gamma_3 = 0$), while 
the other two are broadened and split as two satellite peaks.  
If one notes that the separation $\Delta k$ between 
Dirac cones for the present model with $\gamma_3/t =1.5$ and $\gamma_1/t = 0.2$ is 
order of $0.2a^{-1}$ (Fig.\ref{fig1}(b)), one might 
expect that  there should be a considerable disorder scattering among these Dirac cones, 
since the correlation length  $\eta /a =2$ of disorder is smaller than $\Delta k^{-1} \sim 5a$.
In actual bilayer graphene, where $\gamma_3$ is much smaller, the separation 
$\Delta k$ is estimated to be $\Delta k = 2
\gamma_1 \gamma_3/(3t^2a)$ \cite{MF}, which gives $\Delta k^{-1} \sim 120 a \sim 17$nm for the bulk values of $\gamma_1$ 
and $\gamma_3$. The scale of ripples, on the other hand, is estimated to be $10 \sim 15$nm \cite{Meyer,Geringer}, which 
is again smaller than $\Delta k^{-1}  \sim 17$nm.
We thus expect that the mixing of the four Dirac cones due to ripples can be also relevant in actual bilayer graphene.

\section{Summary and discussions}
We have investigated the stability of zero-mode Landau levels of bilayer graphene in small magnetic fields 
where  the trigonal warping is relevant. We have considered a bond disorder that respects the chiral symmetry 
and is correlated over 
a few lattice constants, which suppresses the inter-valley scattering but causes the intra-valley scattering.
We have found that, among four-fold Landau levels per spin and per valley,
two Landau levels are stable against such a disorder 
with the anomalously sharp density of states at $E=0$. 
On the other hand, the other two levels are broadened and yield split satellite 
peaks. 

This result can be attributed to the fact that the total chirality of 
the four Dirac cones at K (K')  is $2 (-2)$, which is topologically protected even in the presence of 
intra-valley scattering due to disorder. When the inter-valley scattering is 
switched on by making the 
the disorder uncorrelated spatially ($\eta=0$), the anomalously sharp peak is 
washed away (Fig. \ref{fig2}(a)).  
The present results then suggest that the mixing of the two Dirac cones with opposite chirality by the disorder scattering 
generally destroys the anomalous stability of  zero-mode Landau levels of each Dirac cone.

\section*{Acknowledgments}
This work is partly supported by the Grants-in-Aid for Scientific Research (No. 22540336 and No. 23340112) from JSPS.

\end{document}